# A 3D Sweep Hull Algorithm for computing Convex Hulls and Delaunay Triangulation.

Dr David. A. Sinclair
david@newtonapples.net
Imense Ltd

## Abstract

This paper presents a new O(nlog(n)) algorithm for computing the convex hull of a set of 3 dimensional points. The algorithm first sorts the point in (x,y,z) then incrementally adds sorted points to the convex hull using the constraint that each new point added to the hull can 'see' at least one facet touching the last point added. The reduces the search time for adding new points. The algorithm belongs to the family of swept hull algorithms.

While slower than q-hull for the general case it significantly outperforms q-hull for the pathological case where all of the points are on the 3D hull (as is the case for Delaunay triangulation). The algorithm has been named the 'Newton Apple Wrapper algorithm' and has been released under GPL in C++.

keywords: Delaunay triangulation, 3D convex hull.

## Introduction

This paper presents a new 3D convex hull algorithm (named the Newton Apple Wrapper algorithm or 'NAW' algorithm for short) that performs efficiently in the case were all of the points are on the hull. This paper is not intended as an introduction to convex hulls, Delaunay triangulation or computational geometry.

The NAW algorithm is primarily intended for Delaunay triangulation and is benchmarked against q-hull and sweep-line.

## The Newton Apple Wrapper algorithm.

The NAW algorithm functions as follows:

1) Sort a set of 3D point in ascending z (x (y)).
2) Starting with the first of the sorted points add more points until a non-zero area triangle (or triangles) is created to form the seed hull.
3) Sequentially add new points to the hull. The facets of the hull are triangles, these are maintained as a list with adjacency information. The process of adding a new point to the hull involves determining which triangular facets (on the hull) are visible to the new point and replacing them with new triangles made using the new point and the occluding edges of the hull from the perspective of the new point.

The cleverness in the algorithm is in the maintenance of the list of facets and their adjacencies. The sorting of the points in z(x(y)) guarantees that when a new point is to be added to the hull the last point added will be the vertex of at least one 'visible facet' of the existing hull. Convexity of the hull guarantees that all of the facets visible to the new point will be single connected. When searching for the visible facets of the

hull it is sufficient to start by testing the facets which have
the last added point as a vertex until one is found and then
neighbour relations can be used to find the complete set of
visible facets.

Figure 1 illustrates the process of inserting a new points into
the convex hull of an existing set of points. A set of 100
randomly generated points as (x,y,x*x + y*y).

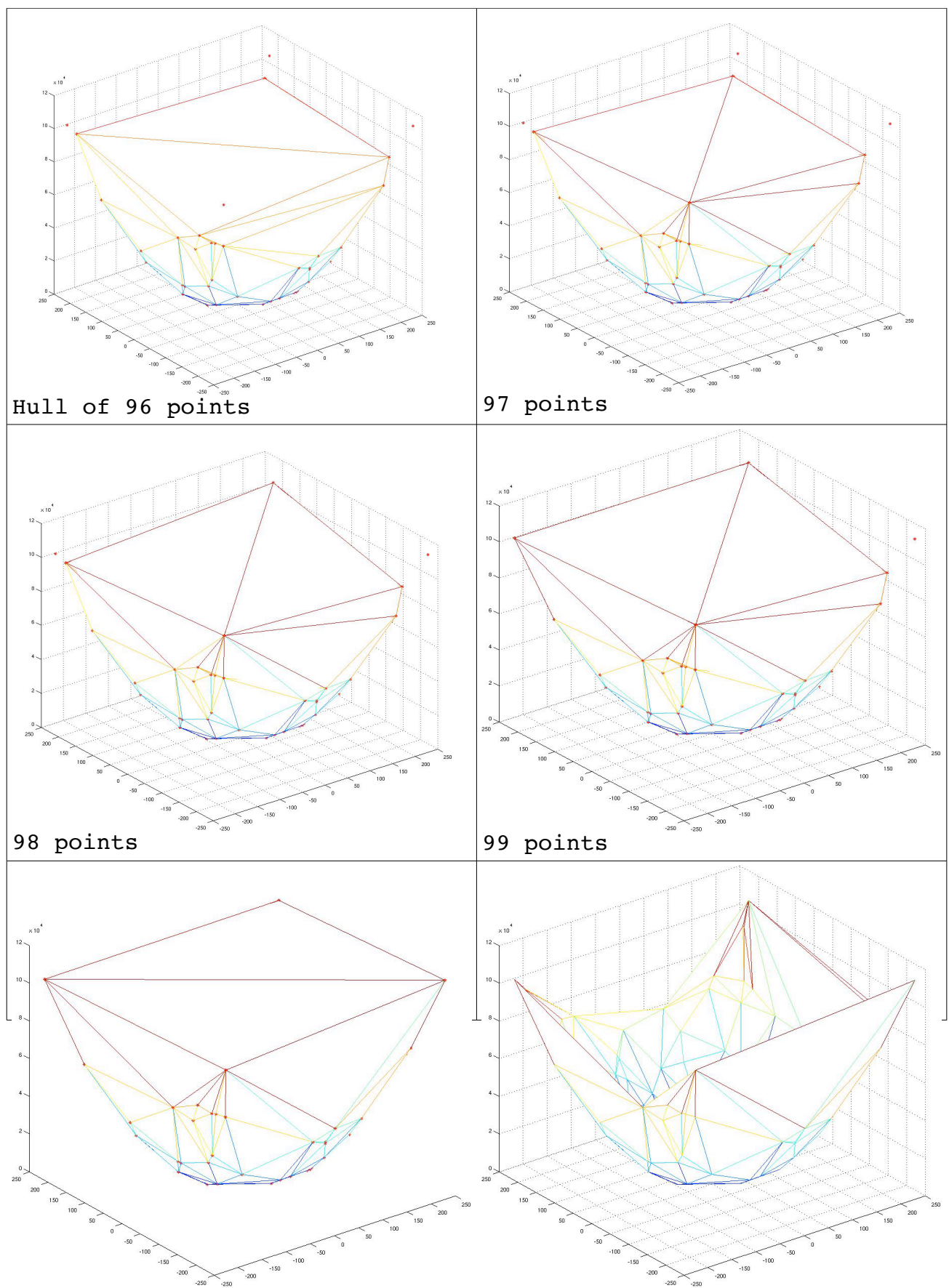

Hull of 96 points | 97 points
98 points | 99 points

| Hull of 100 points | Downwards facing facets only. |
|---|---|
| 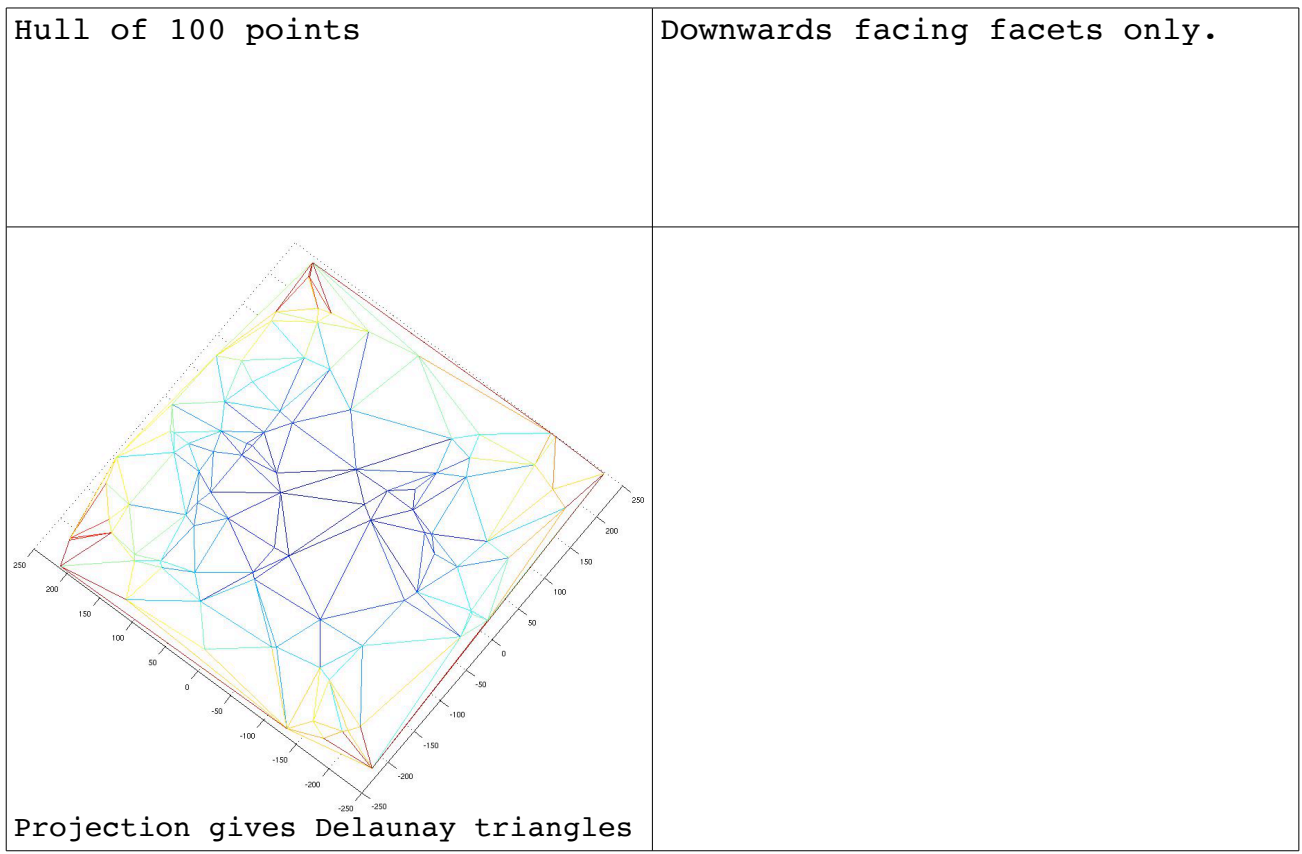 Projection gives Delaunay triangles | |

Figure 1.
*Sequential insertion of randomly generated points on a parabola of revolution.*

## Computational order.

The algorithm is dependent on the receiving a sorted set of 3D points, this bounds the algorithms performance by the time taken to sort the points and this typically will mean using QuickSort meaning the expected performance of the algorithm cannot exceed $O(n\log(n))$. The algorithm is primarily aimed at performing Delaunay triangulation of a set of 2D (*x,y*) points through mapping them into 3D as (*x,y, x*x+y*y* ) i.e. the points are constrained to lie on a parabola of revolution in 3D and ALL of the point will be on the resulting convex hull. Point insertion time is a function of the number of triangular facets visible to each new point. Figure 2 shows a graph of the number of insertions required as points in sets of 1000 and 10,000 points mapped into a hull.

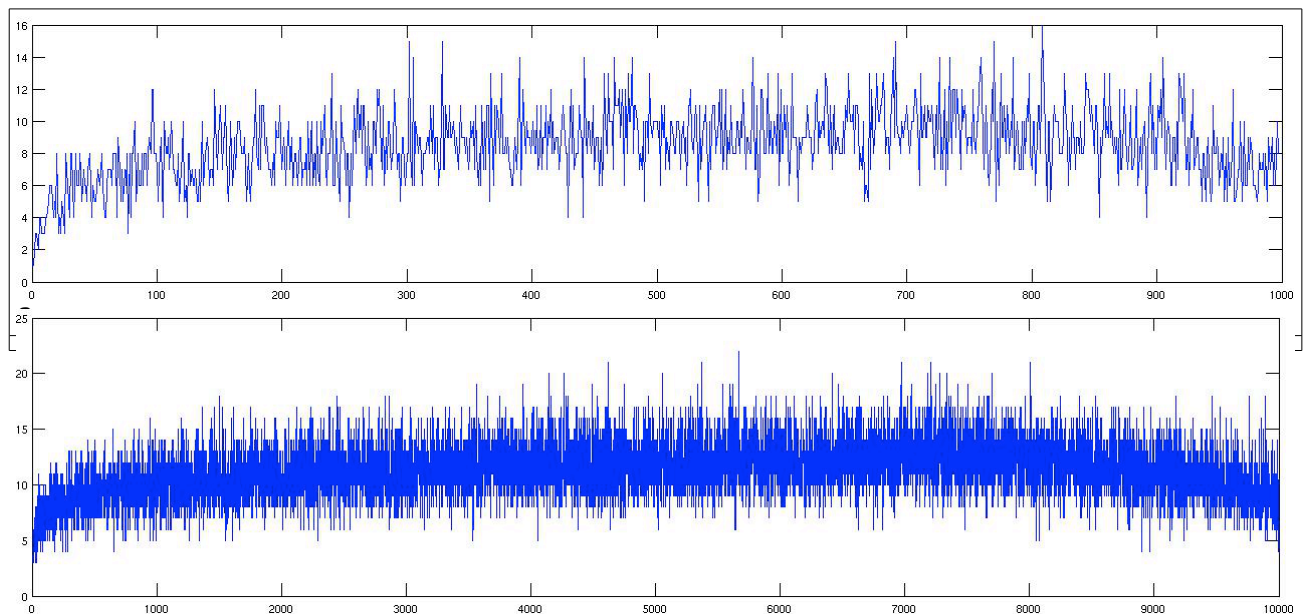

b.

Figure 2.
*Graph of number of visible facets on the hull as points are inserted.*

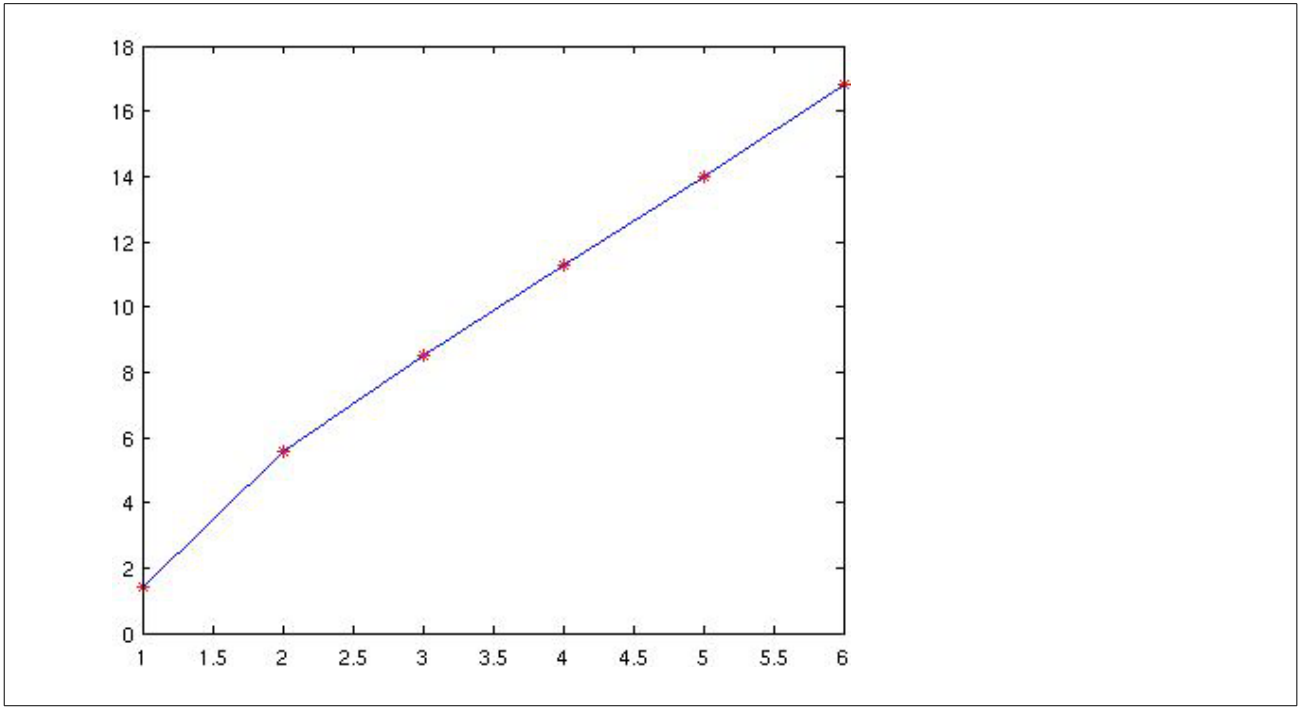

Figure 3 shows a graph of the mean number of visible facets as a function of log of the size of the point set. As can be seen the mean number of visible facets is linear in the exponent of the number of points which demonstrates that the NAW algorithm is O(nlog(n)) in complexity when used for Delaunay triangulation.

**Performance comparison.**

A series of random 2D points were generated with between 100 and 1,000,000 points and Delaunay triangulations computed using the following open source Delaunay triangulation algorithms: Steve Fortune's Sweep-Line algorithm, qhull (from qhull.org), S-HullPro and NAW.

|            | 100 pts   | 1000 pts | 10000 pts | 10^5 pts | 10^6 pts |
|------------|-----------|----------|-----------|----------|----------|
| Sweep-Line | <0.002s   | <0.005s  | 0.043s    | 0.47s    | 6.079s   |
| qhull      | 0.00055s  | 0.00428s | 0.0438s   | 0.85s    | 10.84s   |
| S-HullPro  | 0.00046s  | 0.0037s  | 0.0437s   | 0.619s   | 8.47s    |
| NAW        | 0.001s    | 0.0028s  | 0.0428s   | 0.418s   | 6.104    |

**Summary and conclusions.**

The NAW algorithm is faster than qhull Delaunay triangulation of larger sets of points. It is a conceptually simpler algorithm than Sweep-Line but numerically less stable. The algorithm is $O(n\log(n))$ where n is the number of points. There is very likely room to improve the performance of the code, which is released as GPL in C++.

**References.**

http://www.wikipedia.org, http://google.com search for convex hull 3d or Delaunay triangulation.

**Appendix A.**

Code listing for NAW algorithm.

**NewtonApple_hull3D.h**

```
#ifndef _structures_h
#define _structures_h

// for FILE

#include <stdlib.h>
#include <vector>
#include <set>

/* copyright 2016 Dr David Sinclair
   david@newtonapples.net
   all rights reserved.

   this code is released under GPL3,
   a copy of the license can be found at
   http://www.gnu.org/licenses/gpl-3.0.html

   you can purchase a un-restricted license from
   http://newtonapples.net

   where algorithm details are explained.
```

```
struct Tri
{
  int id, keep;
  int a,b, c;
  int ab, bc, ac;  // adjacent edges index to neighbouring triangle.
  float er, ec, ez; // visible normal to triangular facet.

  Tri() {};
  Tri(int x, int y, int q) : id(0), keep(1),
                             a(x), b(y),c(q),
                             ab(-1), bc(-1), ac(-1),
                             er(0), ec(0), ez(0) {};
  Tri(const Tri &p) : id(p.id), keep(p.keep),
                     a(p.a), b(p.b), c(p.c),
                     ab(p.ab), bc(p.bc), ac(p.ac),
                     er(p.er), ec(p.ec), ez(p.ez) {};

  Tri &operator=(const Tri &p)
  {
    id = p.id;
    keep = p.keep;
    a = p.a;
    b = p.b;
    c = p.c;

    ab = p.ab;
    bc = p.bc;
    ac = p.ac;

    er = p.er;
    ec = p.ec;
    ez = p.ez;

    return *this;
  };
};

struct R3
{
  int id;
  float r,c, z ;
  R3() {};
  R3(float a, float b, float x) : r(a), c(b), z(x), id(-1) {};
  R3(const R3 &p) : id(p.id),
                    r(p.r), c(p.c), z(p.z){};

  R3 &operator=(const R3 &p)
  {
    id = p.id;
    r = p.r;
    c = p.c;
    z = p.z;
    return *this;
  };

};

// sort into descending order (for use in corner responce ranking).
inline bool operator<(const R3 &a, const R3 &b)
{
  if( a.z == b.z){
    if( a.r == b.r ){
      return a.c < b.c;
    }
    return a.r < b.r;
```

```cpp
  }
  return a.z <  b.z;
};

struct Snork
{
  int id;
  int a,b ;
  Snork() : id(-1), a(0), b(0) {};
  Snork(int i, int r, int x) : id(i), a(r), b(x) {};
  Snork(const Snork &p) : id(p.id), a(p.a), b(p.b){};

  Snork &operator=(const Snork &p)
  {
    id = p.id;
    a = p.a;
    b = p.b;

    return *this;
  };

};

// sort into descending order (for use in corner responce ranking).
inline bool operator<(const Snork &a, const Snork &b)
{
  if( a.a == b.a ){
    return a.b < b.b;
  }
  return a.a < b.a;

};

// from NewtonApple_hull3D.cpp

int  read_R3   (std::vector<R3> &pts, char * fname);
void write_R3  (std::vector<R3> &pts, char * fname);
void write_Tris(std::vector<Tri> &ts, char * fname);

int  de_duplicateR3( std::vector<R3> &pts, std::vector<int> &outx,std::vector<R3> &pts2 );

int NewtonApple_Delaunay( std::vector<R3> &pts, std::vector<Tri> &hulk);
int NewtonApple_hull_3D ( std::vector<R3> &pts, std::vector<Tri> &hull);

int  init_hull3D   ( std::vector<R3> &pts, std::vector<Tri> &hull);
void add_coplanar  ( std::vector<R3> &pts, std::vector<Tri> &hull, int id);
int  cross_test    ( std::vector<R3> &pts, int A, int B, int C, int X,
                     float &er, float &ec, float &ez);

#endif
```

# NewtonApple_hull3D.cpp

```cpp
#include <iostream>
//#include <hash_set.h>
//#include <hash_set>
#include <set>
#include <vector>
#include <fstream>
#include <stdlib.h>
#include <math.h>
#include <string>
#include <algorithm>

#include "NewtonApple_hull3D.h"

using namespace std;

/* copyright 2016 Dr David Sinclair
      david@newtonapples.net
      all rights reserved.

      version of 10-feb-2016.

      this code is released under GPL3,
      a copy of the license can be found at
      http://www.gnu.org/licenses/gpl-3.0.html

      you can purchase a un-restricted license from
      http://newtonapples.net

      where algorithm details are explained.

      If you do choose to purchase a license I will do my best
      to post you a free bag of Newton Apple Chocolates!,
      the cleverest chocolates on the Internet.

 */

/*
     read an ascii file of (r,c) or (r,c,z) point pairs.

     the first line of the points file should contain
     "<NUMP>  2 points" or maybe      "<NUMP>   3 points"

     if it does not have the word points in it the first line is
     interpretted as a point pair or triplod.

 */
int read_R3(std::vector<R3> &pts, char * fname){
  char s0[513];
  int nump =0;
  float p1,p2, p3;

  R3 pt;

  std::string line;
  std::string points_str("points");

  std::ifstream myfile;
  myfile.open(fname);

  if (myfile.is_open()){

    getline (myfile,line);
    //int numc = line.length();

    // check string for the string "points"
    int n = (int) line.find( points_str);
    if( n < line.length() ){
      while ( myfile.good() ){
        getline (myfile,line);
```

```cpp
      if( line.length() <= 512){
        copy( line.begin(), line.end(), s0);
        s0[line.length()] = 0;
        int v = sscanf( s0, "%g %g %g", &p1,&p2, &p3);
        if( v == 3 ){
          pt.id = nump;
          nump++;
          pt.r = p1;
          pt.c = p2;
          pt.z = p3;

          pts.push_back(pt);
        }

        else{
          v = sscanf( s0, "%g %g", &p1,&p2);
          if( v == 2 ){
            pt.id = nump;
            nump++;
            pt.r = p1;
            pt.c = p2;
            pt.z = p1*p1 + p2*p2;

            pts.push_back(pt);
          }
        }
      }
    }
  }
  else{   // assume all number pairs on a line are points
    if( line.length() <= 512){
      copy( line.begin(), line.end(), s0);
      s0[line.length()] = 0;
      int v = sscanf( s0, "%g %g %g", &p1,&p2, &p3);
      if( v == 3 ){
        pt.id = nump;
        nump++;
        pt.r = p1;
        pt.c = p2;
        pt.z = p3;

        pts.push_back(pt);
      }

      else{
        v = sscanf( s0, "%g %g", &p1,&p2);
        if( v == 2 ){
          pt.id = nump;
          nump++;
          pt.r = p1;
          pt.c = p2;
          pt.z = p1*p1 + p2*p2;

          pts.push_back(pt);
        }
      }

    }
    while ( myfile.good() ){
     getline (myfile,line);
     if( line.length() <= 512){
        copy( line.begin(), line.end(), s0);
        s0[line.length()] = 0;
        int v = sscanf( s0, "%g %g %g", &p1,&p2, &p3);
        if( v == 3 ){
          pt.id = nump;
          nump++;
          pt.r = p1;
          pt.c = p2;
          pt.z = p3;

          pts.push_back(pt);
        }

        else{
          v = sscanf( s0, "%g %g", &p1,&p2);
          if( v == 2 ){
            pt.id = nump;
```

```cpp
          nump++;
          pt.r = p1;
          pt.c = p2;
          pt.z = p1*p1 + p2*p2;

          pts.push_back(pt);
        }
      }
     }
    }
   }
   myfile.close();
  }

  nump = (int) pts.size();

  return(nump);
};

/*
      write out a set of points to disk

 */

void write_R3(std::vector<R3> &pts, char * fname){
   std::ofstream out(fname, ios::out);

   int nr = (int) pts.size();
   out << nr << " 3 points" << endl;

   for (int r = 0; r < nr; r++){
     out << pts[r].r << ' ' << pts[r].c <<  ' ' << pts[r].z << endl;
   }
   out.close();

   return;
};

/*
 write out triangle ids to be compatible with matlab/octave array numbering.

 */
void write_Tris(std::vector<Tri> &ts, char * fname){
   std::ofstream out(fname, ios::out);

   int nr = (int) ts.size();
   out << nr << " 6   point-ids (1,2,3)  adjacent triangle-ids ( limbs ab  ac  bc )" << endl;

   for (int r = 0; r < nr; r++){
     out << ts[r].a+1 << ' ' << ts[r].b+1 <<' ' << ts[r].c+1 <<' '
       << ts[r].ab+1 <<' ' << ts[r].ac+1 <<' ' << ts[r].bc+1 << endl; //" " << ts[r].ro <<  endl;
   }
   out.close();

   return;
};

/*  To create a Delaunay triangulation from a 2D point set:
    make the 'z' coordinate = (x*x + y*y),
    find the convex hull in R3 of a point cloud.
    using a sweep-hull algorithm called the Newton Apple Wrapper.
    discard the facets that are not downward facing.

 */

int NewtonApple_Delaunay( std::vector<R3> &pts, std::vector<Tri> &hulk)
{

  int nump = (int) pts.size();

  if( nump <= 4 ){
    cerr << "less than 4 points, aborting " << endl;
```

```cpp
      return(-1);
    }

    sort( pts.begin(), pts.end() );

    std::vector<Tri> hull;

    int num = init_hull3D(pts, hull);

    //    return(0); // exit here is you do not need to write the triangles to disk.

    // just pick out the hull triangles and renumber.
    int numh = hull.size();
    int *taken = new int [numh];
    //int taken[numh];
    memset(taken, -1, numh*sizeof(int));

    int cnt = 0;
    for(int t=0; t<numh; t++){  // create an index from old tri-id to new tri-id.
      if( hull[t].keep > 0 ){   // point index remains unchanged.
        taken[t] = cnt;
        cnt ++;
      }
    }

    for(int t=0; t<numh; t++){  // create an index from old tri-id to new tri-id.
      if( hull[t].keep > 0 ){   // point index remains unchanged.
        Tri T = hull[t];
        T.id = taken[t];
        if( taken[T.ab] < 0 ){
         cerr << "broken hull" << endl;
         delete [] taken;
         exit(0);
        }
        T.ab = taken[T.ab];

        if( taken[T.bc] < 0 ){
         cerr << "broken hull" << endl;
         delete [] taken;
         exit(0);
        }
        T.bc = taken[T.bc];

        if( taken[T.ac] < 0 ){
         cerr << "broken hull" << endl;
         delete [] taken;
         exit(0);
        }
        T.ac = taken[T.ac];

        // look at the normal to the triangle
        if( hull[t].ez < 0 ){
         hulk.push_back(T);
        }
      }
    }

    delete [] taken;

  return(1);
}

/*
    find the convex hull in R3 of a point cloud.
    using a sweep-hull algorithm called the Newton Apple Wrapper.

 */

int NewtonApple_hull_3D( std::vector<R3> &pts, std::vector<Tri> &hulk)
{

    int nump = (int) pts.size();
```

```cpp
  if( nump <= 4 ){
    cerr << "less than 4 points, aborting " << endl;
    return(-1);
  }

  sort( pts.begin(), pts.end() );

  std::vector<Tri> hull;

  int num = init_hull3D(pts, hull);

  //   return(0); // exit here is you do not need to write the triangles to disk.

  // just pick out the hull triangles and renumber.
  int numh = hull.size();
  int *taken = new int [numh];
  //int taken[numh];
  memset(taken, -1, numh*sizeof(int));

  int cnt = 0;
  for(int t=0; t<numh; t++){  // create an index from old tri-id to new tri-id.
    if( hull[t].keep > 0 ){   // point index remains unchanged.
      taken[t] = cnt;
      cnt ++;
    }
  }

  for(int t=0; t<numh; t++){  // create an index from old tri-id to new tri-id.
    if( hull[t].keep > 0 ){   // point index remains unchanged.
      Tri T = hull[t];
      T.id = taken[t];
      if( taken[T.ab] < 0 ){
       cerr << "broken hull" << endl;
       delete [] taken;
       exit(0);
      }
      T.ab = taken[T.ab];

      if( taken[T.bc] < 0 ){
       cerr << "broken hull" << endl;
       delete [] taken;
       exit(0);
      }
      T.bc = taken[T.bc];

      if( taken[T.ac] < 0 ){
       cerr << "broken hull" << endl;
       delete [] taken;
       exit(0);
      }
      T.ac = taken[T.ac];

      hulk.push_back(T);
    }
  }

  delete [] taken;

  return(1);
}
// if you are de-duplicating points in R2 remember to set .z to something sensible.

int de_duplicateR3( std::vector<R3> &pts, std::vector<int> &outx,std::vector<R3> &pts2 ){

  int nump = (int) pts.size();
  std::vector<R3> dpx;
  R3 d;
  for( int k=0; k<nump; k++){
    d.r = pts[k].r;
    d.c = pts[k].c;
    d.z = pts[k].z;
    d.id = k;
    dpx.push_back(d);
  }

  sort(dpx.begin(), dpx.end());
```

```
    cerr << "de-duplicating ";
    pts2.clear();
    pts2.push_back(pts[dpx[0].id]);
    pts2[0].id = 0;
    int cnt = 1;

    for( int k=0; k<nump-1; k++){
      if( dpx[k].r == dpx[k+1].r && dpx[k].c == dpx[k+1].c && dpx[k].z == dpx[k+1].z ){
        outx.push_back( dpx[k+1].id);
      }
      else{
        pts[dpx[k+1].id].id = cnt;
        pts2.push_back(pts[dpx[k+1].id]);
        cnt++;
      }
    }

    cerr << "removed  " << outx.size() << " points " << endl;

    return(outx.size());
}

// initialise the hull to the point where there is a not zero volume
// hull.

int init_hull3D( std::vector<R3> &pts, std::vector<Tri> &hull)
{

  int nump = (int) pts.size();

  float mr=0, mc = 0, mz = 0;
  float Mr=0, Mc = 0, Mz = 0;

  Tri T1(0,1,2);
  float r0 = pts[0].r, c0 = pts[0].c, z0 = pts[0].z;
  float r1 = pts[1].r, c1 = pts[1].c, z1 = pts[1].z;
  float r2 = pts[2].r, c2 = pts[2].c, z2 = pts[2].z;

  Mr = r0+r1+r2;
  Mc = c0+c1+c2;
  Mz = z0+z1+z2;

  // check for colinearity
  float r01 = r1-r0, r02 = r2-r0;
  float c01 = c1-c0, c02 = c2-c0;
  float z01 = z1-z0, z02 = z2-z0;

  float e0 = c01*z02 - c02*z01;
  float e1 = -r01*z02 + r02*z01;
  float e2 = r01*c02 - r02*c01;

  if( e0==0 && e1==0 && e2==0 ){ // do not add a facet.
     cerr << "stop fucking me arround and give me a valid opening facet, you tit. " << endl;
     return(-1);
  }
  T1.id = 0;
  T1.er = e0;
  T1.ec = e1;
  T1.ez = e2;

  T1.ab = 1;    // adjacent facet id number
  T1.ac = 1;
  T1.bc = 1;

  hull.push_back( T1);

  T1.id = 1;
  T1.er = -e0;
  T1.ec = -e1;
  T1.ez = -e2;

  T1.ab = 0;
  T1.ac = 0;
  T1.bc = 0;
```

```cpp
    hull.push_back(T1);
    std::vector<int> xlist;
    Tri Tnew;

    int numt = 2;

    for( int p=3; p<nump; p++){ // add points until a non coplanar set of points is achieved.
        R3 &pt = pts[p];

        Mr += pt.r; mr = Mr/(p+1);
        Mc += pt.c; mc = Mc/(p+1);
        Mz += pt.z; mz = Mz/(p+1);

        // find the first visible plane.
        int numh = hull.size();
        int hvis = -1;
        float r = pt.r;
        float c = pt.c;
        float z = pt.z;
        xlist.clear();

        for( int h=numh-1; h>=0; h--){
            Tri &t= hull[h];
            float R1 = pts[t.a].r;
            float C1 = pts[t.a].c;
            float Z1 = pts[t.a].z;

            float dr = r-R1;
            float dc = c-C1;
            float dz = z-Z1;

            float d = dr*t.er + dc*t.ec + dz*t.ez;

            if( d > 0 ){
             hvis = h;
             hull[h].keep = 0;
             xlist.push_back(hvis);
             break;
            }
        }
        if( hvis < 0 ){
            add_coplanar(pts, hull, p);
        }
        if( hvis >= 0 ){
            // new triangular facets are formed from neighbouring invisible planes.
            int numh = hull.size();
            int numx = xlist.size();
            for( int x=0; x<numx; x++){
             int xid = xlist[x];
             int ab = hull[xid].ab;     // facet adjacent to line ab
             Tri &tAB= hull[ab];

             float R1 = pts[tAB.a].r;   // point on next triangle
             float C1 = pts[tAB.a].c;
             float Z1 = pts[tAB.a].z;

             float dr = r-R1;
             float dc = c-C1;
             float dz = z-Z1;

             float d = dr*tAB.er + dc*tAB.ec + dz*tAB.ez;

             if( d > 0 ){ // add to xlist.
                if( hull[ab].keep == 1){
                    hull[ab].keep = 0;
                    xlist.push_back(ab);
                    numx ++;
                }
             }
             else{ // spawn a new triangle.
                Tnew.id = hull.size();
                Tnew.keep = 2;
                Tnew.a = p;
                Tnew.b = hull[xid].a;
                Tnew.c = hull[xid].b;

                Tnew.ab = -1;
                Tnew.ac = -1;
```

```cpp
    Tnew.bc = ab;

    // make normal vector.
    float dr1 = pts[Tnew.a].r- pts[Tnew.b].r, dr2 = pts[Tnew.a].r- pts[Tnew.c].r;
    float dc1 = pts[Tnew.a].c- pts[Tnew.b].c, dc2 = pts[Tnew.a].c- pts[Tnew.c].c;
    float dz1 = pts[Tnew.a].z- pts[Tnew.b].z, dz2 = pts[Tnew.a].z- pts[Tnew.c].z;

    float er = (dc1*dz2-dc2*dz1);
    float ec = -(dr1*dz2-dr2*dz1);
    float ez = (dr1*dc2-dr2*dc1);

    dr = mr-r; // points from new facet towards [mr,mc,mz]
    dc = mc-c;
    dz = mz-z;
    // make it point outwards.

    float dromadery = dr*er +  dc*ec + dz*ez;

    if( dromadery > 0 ){
      Tnew.er = -er;
      Tnew.ec = -ec;
      Tnew.ez = -ez;
    }
    else{
      Tnew.er = er;
      Tnew.ec = ec;
      Tnew.ez = ez;
    }

    // update the touching triangle tAB
    int A = hull[xid].a, B = hull[xid].b;
    if( (tAB.a == A && tAB.b == B ) ||(tAB.a == B && tAB.b == A ) ){
      tAB.ab = hull.size();
    }
    else if( (tAB.a == A && tAB.c == B ) ||(tAB.a == B && tAB.c == A ) ){
      tAB.ac = hull.size();
    }
    else if( (tAB.b == A && tAB.c == B ) ||(tAB.b == B && tAB.c == A ) ){
      tAB.bc = hull.size();
    }
    else{
      cerr << "Oh crap, the di-lithium crystals are fucked!" << endl;
      cerr << "numeric stability issue, exiting now." << endl;
      exit(9);
    }

    hull.push_back(Tnew);

  }

  // second side of the struck out triangle

  int ac = hull[xid].ac;      // facet adjacent to line ac
  Tri &tAC = hull[ac];

  R1 = pts[tAC.a].r;  // point on next triangle
  C1 = pts[tAC.a].c;
  Z1 = pts[tAC.a].z;

  dr = r-R1;
  dc = c-C1;
  dz = z-Z1;

  d = dr*tAC.er + dc*tAC.ec + dz*tAC.ez;

  if( d > 0 ){ // add to xlist.
    if( hull[ac].keep == 1){
      hull[ac].keep = 0;
      xlist.push_back(ac);
      numx ++;
    }
  }
  else{ // spawn a new triangle.
    Tnew.id = hull.size();
    Tnew.keep = 2;
    Tnew.a = p;
    Tnew.b = hull[xid].a;
    Tnew.c = hull[xid].c;
```

```cpp
      Tnew.ab = -1;
      Tnew.ac = -1;
      Tnew.bc = ac;

      // make normal vector.
      float dr1 = pts[Tnew.a].r- pts[Tnew.b].r, dr2 = pts[Tnew.a].r- pts[Tnew.c].r;
      float dc1 = pts[Tnew.a].c- pts[Tnew.b].c, dc2 = pts[Tnew.a].c- pts[Tnew.c].c;
      float dz1 = pts[Tnew.a].z- pts[Tnew.b].z, dz2 = pts[Tnew.a].z- pts[Tnew.c].z;

      float er =  (dc1*dz2-dc2*dz1);
      float ec = -(dr1*dz2-dr2*dz1);
      float ez =  (dr1*dc2-dr2*dc1);

      dr = mr-r; // points from new facet towards [mr,mc,mz]
      dc = mc-c;
      dz = mz-z;
      // make it point outwards.

      float dromadery = dr*er +  dc*ec + dz*ez;

      if( dromadery > 0 ){
        Tnew.er = -er;
        Tnew.ec = -ec;
        Tnew.ez = -ez;
      }
      else{
        Tnew.er = er;
        Tnew.ec = ec;
        Tnew.ez = ez;
      }
      // update the touching triangle tAC
      int A = hull[xid].a, C = hull[xid].c;
      if( (tAC.a == A && tAC.b == C ) ||(tAC.a == C && tAC.b == A ) ){
        tAC.ab = hull.size();
      }
      else if( (tAC.a == A && tAC.c == C ) ||(tAC.a == C && tAC.c == A ) ){
        tAC.ac = hull.size();
      }
      else if( (tAC.b == A && tAC.c == C ) ||(tAC.b == C && tAC.c == A ) ){
        tAC.bc = hull.size();
      }
      else{
        cerr << "Oh crap, warp drive failure, dude!" << endl;
        cerr << "numeric stability issue, exiting now." << endl;
        exit(9);
      }

      hull.push_back(Tnew);
    }

    // third side of the struck out triangle

    int bc = hull[xid].bc;      // facet adjacent to line ac
    Tri &tBC = hull[bc];

    R1 = pts[tBC.a].r;   // point on next triangle
    C1 = pts[tBC.a].c;
    Z1 = pts[tBC.a].z;

    dr = r-R1;
    dc = c-C1;
    dz = z-Z1;

    d = dr*tBC.er + dc*tBC.ec + dz*tBC.ez;

    if( d > 0 ){ // add to xlist.
      if( hull[bc].keep == 1){
        hull[bc].keep = 0;
        xlist.push_back(bc);
        numx ++;
      }
    }
    else{ // spawn a new triangle.
      Tnew.id = hull.size();
      Tnew.keep = 2;
```

```
    Tnew.a = p;
    Tnew.b = hull[xid].b;
    Tnew.c = hull[xid].c;

    Tnew.ab = -1;
    Tnew.ac = -1;
    Tnew.bc = bc;

    // make normal vector.
    float dr1 = pts[Tnew.a].r- pts[Tnew.b].r, dr2 = pts[Tnew.a].r- pts[Tnew.c].r;
    float dc1 = pts[Tnew.a].c- pts[Tnew.b].c, dc2 = pts[Tnew.a].c- pts[Tnew.c].c;
    float dz1 = pts[Tnew.a].z- pts[Tnew.b].z, dz2 = pts[Tnew.a].z- pts[Tnew.c].z;

    float er = (dc1*dz2-dc2*dz1);
    float ec = -(dr1*dz2-dr2*dz1);
    float ez = (dr1*dc2-dr2*dc1);

    dr = mr-r; // points from new facet towards [mr,mc,mz]
    dc = mc-c;
    dz = mz-z;
    // make it point outwards.

    float dromadery = dr*er +  dc*ec + dz*ez;

    if( dromadery > 0 ){
      Tnew.er = -er;
      Tnew.ec = -ec;
      Tnew.ez = -ez;
    }
    else{
      Tnew.er = er;
      Tnew.ec = ec;
      Tnew.ez = ez;
    }

    // update the touching triangle tBC
    int B = hull[xid].b, C = hull[xid].c;
    if( (tBC.a == B && tBC.b == C ) ||(tBC.a == C && tBC.b == B ) ){
      tBC.ab = hull.size();
    }
    else if( (tBC.a == B && tBC.c == C ) ||(tBC.a == C && tBC.c == B ) ){
      tBC.ac = hull.size();
    }
    else if( (tBC.b == B && tBC.c == C ) ||(tBC.b == C && tBC.c == B ) ){
      tBC.bc = hull.size();
    }
    else{
      cerr << "Oh crap, rocket engine failure" << endl;
      cerr << "numeric stability issue, exiting now." << endl;
      exit(9);
    }

    hull.push_back(Tnew);
  }

}

// patch up the new triangles in hull.

int numN = hull.size();
std::vector<Snork> norts;
Snork snort;
for( int q = numN-1; q>= numh; q--){
 if( hull[q].keep > 1){
    snort.id = q;
    snort.a = hull[q].b;
    snort.b = 1;
    norts.push_back(snort);

    snort.a = hull[q].c;
    snort.b = 0;
    norts.push_back(snort);

    hull[q].keep = 1;

 }
}
sort( norts.begin(), norts.end());
int nums = norts.size();
```

```cpp
      if( nums >= 2 ){
       for( int s=0; s<nums-1; s++){
          if( norts[s].a == norts[s+1].a ){
            // link triangle sides.
            if( norts[s].b == 1){
              hull[norts[s].id].ab = norts[s+1].id;
            }
            else{
              hull[norts[s].id].ac = norts[s+1].id;
            }

            if( norts[s+1].b == 1){
              hull[norts[s+1].id].ab = norts[s].id;
            }
            else{
              hull[norts[s+1].id].ac = norts[s].id;
            }
          }
       }
      }

    }
    /*else{
       cerr << "still in the coplanar state you fucking baboon..." << endl;
       // rather complicated and need to add points to the 2D-hull as two faced triangles.
       exit(0);
       }*/

  }

  return(0);
}
// add a point coplanar to the existing planar hull in 3D
// this is not an efficient routine and should only be used
// to add in duff (coplanar) pts at the start.
// it should not be called in doing a Delaunay triangulation
// of a 2D set raised into 3D.

void add_coplanar( std::vector<R3> &pts, std::vector<Tri> &hull, int id)
{

  int numh = hull.size();
  float er, ec, ez;
  for( int k=0; k<numh; k++){
    //find vizible edges. from external edges.

    if( hull[k].c == hull[hull[k].ab].c ){ // ->  ab is an external edge.
       // test this edge for visibility from new point pts[id].
       int A = hull[k].a;
       int B = hull[k].b;
       int C = hull[k].c;

       int zot = cross_test( pts, A, B, C, id, er,ec,ez);

       if( zot < 0 ){ // visible edge facet, create 2 new hull plates.
        Tri up, down;
        up.keep = 2;
        up.id = hull.size();
        up.a = id;
        up.b = A;
        up.c = B;

        up.er = er; up.ec = ec; up.ez = ez;
        up.ab = -1; up.ac = -1;

        down.keep = 2;
        down.id = hull.size()+1;
        down.a = id;
        down.b = A;
        down.c = B;

        down.ab = -1; down.ac = -1;
        down.er = -er; down.ec = -ec; down.ez = -ez;

        float xx = hull[k].er*er + hull[k].ec*ec + hull[k].ez*ez;
```

```
      if( xx > 0 ){
         up.bc = k;
         down.bc = hull[k].ab;

         hull[k].ab = up.id;
         hull[down.bc].ab = down.id;
      }
      else{
         down.bc = k;
         up.bc = hull[k].ab;

         hull[k].ab = down.id;
         hull[up.bc].ab = up.id;
      }

      hull.push_back(up);
      hull.push_back(down);
      }
}

   if( hull[k].a == hull[hull[k].bc].a ){     // bc is an external edge.
      // test this edge for visibility from new point pts[id].
      int A = hull[k].b;
      int B = hull[k].c;
      int C = hull[k].a;

      int zot = cross_test( pts, A, B, C, id, er,er,ez);

      if( zot < 0 ){ // visible edge facet, create 2 new hull plates.
      Tri up, down;
      up.keep = 2;
      up.id = hull.size();
      up.a = id;
      up.b = A;
      up.c = B;

      up.er = er; up.ec = ec; up.ez = ez;
      up.ab = -1; up.ac = -1;

      down.keep = 2;
      down.id = hull.size()+1;
      down.a = id;
      down.b = A;
      down.c = B;

      down.ab = -1; down.ac = -1;
      down.er = -er; down.ec = -ec; down.ez = -ez;

      float xx = hull[k].er*er + hull[k].ec*ec + hull[k].ez*ez;
      if( xx > 0 ){
         up.bc = k;
         down.bc = hull[k].bc;

         hull[k].bc = up.id;
         hull[down.bc].bc = down.id;
      }
      else{
         down.bc = k;
         up.bc = hull[k].bc;

         hull[k].bc = down.id;
         hull[up.bc].bc = up.id;
      }

      hull.push_back(up);
      hull.push_back(down);
      }
}

   if( hull[k].b == hull[hull[k].ac].b ){     // ac is an external edge.
      // test this edge for visibility from new point pts[id].
      int A = hull[k].a;
      int B = hull[k].c;
      int C = hull[k].b;
```

```
      int zot = cross_test( pts, A, B, C, id, er,er,ez);

      if( zot < 0 ){ // visible edge facet, create 2 new hull plates.
        Tri up, down;
        up.keep = 2;
        up.id = hull.size();
        up.a = id;
        up.b = A;
        up.c = B;

        up.er = er; up.ec = ec; up.ez = ez;
        up.ab = -1; up.ac = -1;

        down.keep = 2;
        down.id = hull.size()+1;
        down.a = id;
        down.b = A;
        down.c = B;

        down.ab = -1; down.ac = -1;
        down.er = -er; down.ec = -ec; down.ez = -ez;

        float xx = hull[k].er*er + hull[k].ec*ec + hull[k].ez*ez;
        if( xx > 0 ){
          up.bc = k;
          down.bc = hull[k].ac;

          hull[k].ac = up.id;
          hull[down.bc].ac = down.id;
        }
        else{
          down.bc = k;
          up.bc = hull[k].ac;

          hull[k].ac = down.id;
          hull[up.bc].ac = up.id;
        }

        hull.push_back(up);
        hull.push_back(down);
      }
    }

  }

  // fix up the non asigned hull adjecencies (correctly).

  int numN = hull.size();
  std::vector<Snork> norts;
  Snork snort;
  for( int q = numN-1; q>= numh; q--){
    if( hull[q].keep > 1){
      snort.id = q;
      snort.a = hull[q].b;
      snort.b = 1;
      norts.push_back(snort);

      snort.a = hull[q].c;
      snort.b = 0;
      norts.push_back(snort);

      hull[q].keep = 1;

    }
  }

  sort( norts.begin(), norts.end());
  int nums = norts.size();
  Snork snor;
  snor.id = -1; snor.a=-1; snor.b=-1;
  norts.push_back(snor);
  snor.id = -2; snor.a=-2; snor.b=-2;
  norts.push_back(snor);

  if( nums >= 2 ){
    for( int s=0; s<nums-1; s++){
      if( norts[s].a == norts[s+1].a ){
```

```
      // link triangle sides.
      if( norts[s].a != norts[s+2].a ){ // edge of figure case
        if( norts[s].b == 1){
           hull[norts[s].id].ab = norts[s+1].id;
        }
        else{
           hull[norts[s].id].ac = norts[s+1].id;
        }

        if( norts[s+1].b == 1){
           hull[norts[s+1].id].ab = norts[s].id;
        }
        else{
           hull[norts[s+1].id].ac = norts[s].id;
        }
        s++;
      }
      else{ // internal figure boundary 4 junction case.
        int s1 = s+1, s2 = s+2, s3 = s+3;
        int id = norts[s].id;
        int id1 = norts[s1].id;
        int id2 = norts[s2].id;
        int id3 = norts[s3].id;

        // check normal directions of id and id1..3
        float barf = hull[id].er*hull[id1].er + hull[id].ec*hull[id1].ec + hull[id].ez*hull[id1].ez;
        if( barf > 0){
        }
        else{
           barf = hull[id].er*hull[id2].er + hull[id].ec*hull[id2].ec + hull[id].ez*hull[id2].ez;
           if( barf > 0 ){
              int tmp = id2; id2 = id1; id1 = tmp;
              tmp = s2; s2 = s1; s1 = tmp;
           }
           else{
              barf = hull[id].er*hull[id3].er + hull[id].ec*hull[id3].ec + hull[id].ez*hull[id3].ez;
              if( barf > 0 ){
               int tmp = id3; id3 = id1; id1 = tmp;
               tmp = s3; s3 = s1; s1 = tmp;
              }
           }
        }

        if( norts[s].b == 1){
           hull[norts[s].id].ab = norts[s1].id;
        }
        else{
           hull[norts[s].id].ac = norts[s1].id;
        }

        if( norts[s1].b == 1){
           hull[norts[s1].id].ab = norts[s].id;
        }
        else{
           hull[norts[s1].id].ac = norts[s].id;
        }

        // use s2 and s3

        if( norts[s2].b == 1){
           hull[norts[s2].id].ab = norts[s3].id;
        }
        else{
           hull[norts[s2].id].ac = norts[s3].id;
        }

        if( norts[s3].b == 1){
           hull[norts[s3].id].ab = norts[s2].id;
        }
        else{
           hull[norts[s3].id].ac = norts[s2].id;
        }

        s+=3;
      }
    }
```

```cpp
    }
  }

  return;
}
// cross product relative sign test.
// remmebers the cross product of (ab x cx)

int cross_test( std::vector<R3> &pts, int A, int B, int C, int X,
                float &er, float &ec, float &ez)
{

  float Ar = pts[A].r;
  float Ac = pts[A].c;
  float Az = pts[A].z;

  float Br = pts[B].r;
  float Bc = pts[B].c;
  float Bz = pts[B].z;

  float Cr = pts[C].r;
  float Cc = pts[C].c;
  float Cz = pts[C].z;

  float Xr = pts[X].r;
  float Xc = pts[X].c;
  float Xz = pts[X].z;

  float ABr = Br-Ar;
  float ABc = Bc-Ac;
  float ABz = Bz-Az;

  float ACr = Cr-Ar;
  float ACc = Cc-Ac;
  float ACz = Cz-Az;

  float AXr = Xr-Ar;
  float AXc = Xc-Ac;
  float AXz = Xz-Az;

  er =  (ABc*AXz-ABz*AXc);
  ec = -(ABr*AXz-ABz*AXr);
  ez =  (ABr*AXc-ABc*AXr);

  float kr =  (ABc*ACz-ABz*ACc);
  float kc = -(ABr*ACz-ABz*ACr);
  float kz =  (ABr*ACc-ABc*ACr);

  //   look at sign of (ab x ac).(ab x ax)

  float globit =   kr * er +  kc * ec + kz * ez;

  if( globit >  0 ) return(1);
  if( globit == 0 ) return(0);

  return(-1);

}
```

# Makefile

```makefile
# Makefile  for line NAW routine.

LIBDIR = /usr/lib
INCDIR = /usr/include
STLDIR = /usr/include

#CFLAGS = -g  -I$(INCDIR) -I$(STLDIR) -L$(LIBDIR) -Wno-deprecated
#
CFLAGS = -O3 -I$(INCDIR) -I$(STLDIR) -L$(LIBDIR) -Wno-deprecated
LIBS    = -lm
OBJS  = control.o NewtonApple_hull3D.o
HEADS = NewtonApple_hull3D.h

CC = gcc

# Building rules for Makefile

NewtAppWrap3D: $(OBJS) $(HEADS)
	g++ $(CFLAGS)   -o $@ $(OBJS)   $(LIBS)

%.o: %.cpp
	g++ $(CFLAGS) -c  $< -o $@

clean:
	-rm *.o NewtAppWrap3D

control.o:    control.cpp $(HEADS)

NewtonApple_hull3D.o:     NewtonApple_hull3D.cpp $(HEADS)
```

# Control.cpp

```cpp
#include <iostream>
//#include <hash_set>
#include <set>
#include <vector>
#include <fstream>
#include <stdlib.h>
#include <math.h>

#include <memory.h>
//#include <malloc.h>
#include <ctype.h>
#include <string.h>
#include <string>
#include <strings.h>
#include <sys/types.h>

#include "NewtonApple_hull3D.h"

/* copyright 2016 Dr David Sinclair
   david@newtonapples.net
   all rights reserved.

   this code is released under GPL3,
   a copy of the license can be found at
   http://www.gnu.org/licenses/gpl-3.0.html

   you can purchase a un-restricted license from
   http://newtonapples.net

   where algorithm details are explained.

   If you do choose to purchase a license I will do my best
   to post you a free bag of Newton Apple Chocolates,
   the cleverest chocolates on the Internet.

 */

#include <sys/time.h>

using namespace std;

int main(int argc, char *argv[])
{

    if( argc == 1 ){
        cerr << "NewtonAppleWrapper Delaunay triangulation demo" << endl;
        cerr << "usage: />    shullpro3D <points_file> <triangles_file> " << endl;

        float goat =  ( 2147483648.0-1) /100.0;

        std::vector<R3> pts, pts2;
        R3 pt;
        srandom(1);
        pts.clear();

        //       for(int v=0; v<100000; v++){
        //
        for(int v=0; v<100; v++){
         // for(int v=0; v<1000000; v++){
         //for(int v=0; v<2000000; v++){

          pt.id = v;
          pt.r = (((float) rand() / RAND_MAX ) * 500)-250; // pts.txt
          pt.c = (((float) rand() / RAND_MAX ) * 500)-250;
          //pt.r = (((float) rand() / RAND_MAX ) * 50000)-20000; // pts.txt
          //pt.c = (((float) rand() / RAND_MAX ) * 50000)-20000;

          //pt.z = (int)(((float) rand() / RAND_MAX ) * 500);   // for 3D points in box.
```

```cpp
      pt.z = pt.r*pt.r + pt.c*pt.c;    // for beautiful delaunay triangulations.
      pts.push_back(pt);
     }

      std::vector<Tri> tris;

      std::vector<int> outx;
      int nx = de_duplicateR3( pts, outx, pts2);
      pts.clear();

      write_R3(pts2, "pts.mat");
      cerr << pts2.size() << " randomly generated points int R2/R3 written to pts.mat" << endl;

      struct timeval tv1, tv2; // slight swizzle as pt set is now sorted.
      gettimeofday(&tv1, NULL);

      //int ts = NewtonApple_Delaunay( pts2, tris);
      int ts = NewtonApple_hull_3D( pts2, tris);

      gettimeofday(&tv2, NULL);
      float tx =   (tv2.tv_sec + tv2.tv_usec / 1000000.0) - ( tv1.tv_sec + tv1.tv_usec / 1000000.0);
      pts2.clear();

      cerr <<  tx << " seconds for triangulation" << endl;

      write_Tris(tris, "triangles.mat");
      cerr << tris.size() << " triangles written to triangles.mat" << endl;

      exit(0);
    }
    else if( argc > 1){

      cerr << "reading points from " << argv[1] << endl;
      std::vector<R3> pts4, pts3;

      int nump = read_R3(pts4, argv[1]);
      cerr << nump << " points read" << endl;

      // check for duplicates.
      std::vector<int> dupes;
      int numd = de_duplicateR3( pts4, dupes, pts3);
      cerr << "duplicates filtered   " << numd << endl;
      cerr << "writing point ro pts.mat " << endl;

      write_R3(pts3, "pts.mat");

      struct timeval tv1, tv2;
      gettimeofday(&tv1, NULL);

      std::vector<Tri> tris;
      //
     int ts = NewtonApple_Delaunay( pts3, tris);
     //     int ts = NewtonApple_hull_3D( pts3, tris);

      gettimeofday(&tv2, NULL);
      float tx =   (tv2.tv_sec + tv2.tv_usec / 1000000.0) - ( tv1.tv_sec + tv1.tv_usec / 1000000.0);

      cerr <<  tx << " seconds for triangulation" << endl;

      if( argc == 2 ){
        write_Tris(tris, "triangles.mat");
        cerr << tris.size() << " triangles written to triangles.mat" << endl;
      }
      else{
          write_Tris(tris, argv[2]);
        cerr << tris.size() << " triangles written to " << argv[2] << endl;
      }
    }
```

```
    exit(0);
}

#include <sys/time.h>
//static inline
double getpropertime()
{
    struct timeval tv;
    gettimeofday(&tv, NULL);
    return tv.tv_sec + tv.tv_usec / 1000000.0;
}
```